\documentstyle[11pt,epsfig,epsf,axodraw]{article} 
\newcommand{\be}{\begin{equation}}
\newcommand{\ee}{\end{equation}}
\newcommand{\bea}{\begin{eqnarray}}
\newcommand{\eea}{\end{eqnarray}}
\def\bbbar {B^0-\bar{B^0}}
\def\b {B^0}
\def\bbar {\bar{B^0}}

\begin{document}
\thispagestyle{empty}
\title{
\rightline {\small{HIP-2002-61/TH}}\vspace*{-3mm} 
\rightline {\small{DO-TH 02/20}}\vspace*{-3mm} 
\rightline {\small{hep-ph/0212047}}\vspace*{5mm}
\bf
Effects of Universal Extra Dimensions on\\
\boldmath$\bbbar~~$\unboldmath Mixing}
\author{{\large\bf Debrupa Chakraverty} \thanks{E-mail: chakrave@pcu.helsinki.fi},
{\large\bf Katri Huitu}    \thanks{E-mail: huitu@pcu.helsinki.fi}, \\
{\em Helsinki Institute of Physics, P.O. Box 64, 
00014 University of Helsinki, Finland}
\and
{\large\bf Anirban Kundu} \thanks{E-mail: kundu@zylon.physik.uni-dortmund.de}
\thanks{On leave from Department of Physics, Jadavpur University,
Kolkata 700032, India}\\
{\em Universit\"at Dortmund, Institut f\"ur Physik, D-44221 Dortmund, Germany}}

\date{\today}
\maketitle

\begin{abstract}
We study contributions coming to $\Delta m_B$ from one or more universal
extra dimensions (UED) in which all the Standard Model fields can propagate. 
In the model with a single UED, the box diagrams for mixing are convergent
and therefore insensitive to the cutoff scale of the theory. In the case 
of two UEDs, the  result is not very sensitive to the cut-off scale due to 
 GIM mechanism.
 Within the present range of 
the parameters at $1 \sigma$ level, 
the lower bound on the compactification scale $1/R$ has been 
estimated to be 165 GeV for one UED and 280 GeV for two UEDs. The bound 
increases drastically if one can have a better determination of the $B$ 
meson decay constant $f_B$ and the QCD correction parameter $B_B$. For 
example, it rises to 740 GeV (for one UED) if the error (at $1\sigma$)
in the determination 
of $f_B\sqrt{B_B}$ from quenched lattice calculation is reduced to one-third 
from its present value. 
The UED contributions to the $K$ system are strongly suppressed.

\end{abstract}

\noindent PACS number(s):
11.25.Mj, 12.15.Ff, 14.40.Nd
\newpage

Inspired by the string theories, a possible solution to the hierarchy
problem of the Standard Model (SM) may be the higher dimensional 
scenarios. These scenarios get additional motivation from the 
potential to solve some of the open problems of the SM:
gauge coupling unification \cite{gauge}, supersymmetry breaking \cite{susy}, 
neutrino mass generation \cite{neutrino}, and the
explanation of fermion mass hierarchies \cite{fermion}. 
The observation of four dimensional world in our everyday life ensures that 
the extra dimensions are compactified. The simplest way is to
compactify them on a circle of a radius $R$ ($S^1$). A nice feature of 
these theories is that the compactification radius can be large so that
$1/R$ can be as low as a few hundreds of GeV \cite{susy,lowscale}. 
One might argue that in the most natural framework all the
SM fields should be allowed to propagate in the extra dimensions.
Care, however, must be taken to obtain chiral fermions in four dimensions (4D)
from such universal extra dimension (UED) models.

In this letter we confine ourselves to the UED model formulated in \cite{zbb}. 
In 4D effective theory, the existence of these extra 
dimensions are felt by the appearance of towers of heavy Kaluza-Klein (KK) 
states having masses quantised in units of the compactification scale $1/R$. 
The key feature of UED models is  the conservation of momentum in extra 
dimensions which leads to KK number conservation in the effective 4D theory.
Such theories naturally lead to the existence of a lightest KK particle which
is a viable dark matter candidate \cite{servant}. 
One should note in this context that orbifolding, which is necessary to
forbid wrong-chirality fermions at the lowest level, generates 
KK number violating interactions through boundary terms \cite{cms1,georgi}.
Though these interactions have interesting phenomenological implications in the 
decays of such KK modes, we put them to be equal to zero by hand
and do not consider them any further in our study of the virtual 
effects of those modes. 
Consequently in our calculations, there are no
vertices which violate the KK number conservation.
This forbids production of isolated KK particles at colliders and 
tree level contributions to the electroweak observables. 
In the non-universal case, where the fermions (and maybe some of the bosons)
are confined to a 4D brane, the presence of a localising delta function in the 
Lagrangian permits KK number violating couplings, which is not true for the
UED models, and hence none of the existing bounds on non-universal extra 
dimensional models from single KK productions at colliders \cite{collider1}
and from tree as well as loop level electroweak  constraints \cite{loop1} 
are applicable for UEDs.  In UED,
apart from the direct KK pair production at colliders \cite{collider2}, 
one may get indirect bound on
the compactification scale $1/R$ from the virtual effects 
of KK modes at loop level \cite{zbb,loop2,higgs,bsg}.
It is natural to look on processes which are sensitive to 
radiative corrections even in the
absence of KK modes in order to study the dominant 
loop effects induced by the
exchange of them. In the SM, the most important loop 
effects are those enhanced by the heavy top quark mass.
Thus one may get valuable information on 
the size of the extra dimension
through the one loop KK mode contributions 
to the processes,  $Z \to b {\bar b}$ \cite{zbb}, $b
\to s \gamma$ \cite{bsg} and $\bbbar$ mixing.
The lower bound on the compactification scale ($1/R$) from 
collider phenomenology \cite{collider2}, Higgs physics \cite{higgs}, electroweak precision
measurements \cite{zbb}, and flavour changing process $b \to s \gamma$ \cite{bsg}
has been estimated
to lie between 200 and 500 GeV. 

In this letter, we mainly address  the effects of only one spatial UED at
the  one-loop level to $\bbbar$ mixing. The bound on the 
compactification scale is derived taking into account all input
uncertainties (and we also show what happens if these uncertainties come
down) and hence can be regarded as a robust one. A brief qualitative discussion
for more than one extra dimension is also presented, and we comment on the
$K$ system too. As a starting point, we consider the
relevant part of the five dimensional ($5D$) Lagrangian:
\be
{\cal L}_5(x,y) =  i {\bar Q} \Gamma^M D_M Q 
+ \lambda^U_5{\bar Q} U (i\sigma_2 H^*)
 + \lambda_5^D {\bar Q} D H + h.c.
\ee
Here $M$ (1 to 5) is the Lorentz index. The covariant derivative $D_M$ 
can be expressed as
\be
D_M = \partial_M + i \sum_{i=1}^3 g_5^i T_i^a A_{iM}^a,
\ee
where $g_5^i$s are the $5D$ coupling constants 
associated with the SM gauge group $SU(3)_c
\times SU(2)_L\times U(1)_Y$, and $T_i^a$s are the 
corresponding generators.
The parameters
 $\lambda_5^U$ and $\lambda_5^D$ are the 5D Yukawa couplings. The 5D
Dirac matrices are $\Gamma^M \equiv (\gamma^\mu,i\gamma^5)$ 
(see, {\em e.g.}, \cite{proeyen}) and
$y$ denotes the coordinate along the extra dimension. The fields $Q$, $U$ and 
$D$, all functions of $x^\mu$ and $y$,
 describe 5D generic quark doublet, up type quark singlet, and down type
quark singlet, respectively. Unlike in the Standard Model, they have both 
chiralities, and are of vector type.  The field $H$ is the  5D Higgs doublet, 
and the generic 5D gauge bosons for each gauge group are denoted by $A_{iM}^a$.
The component of the gauge bosons along the extra dimension is the pseudoscalar
$A_{i5}$. In order to derive the 4D  Lagrangian we must expand 
the five dimensional fields into their KK modes. To project out the zero modes 
of the wrong chirality  ({\em i.e.}, $Q_R$, $U_L$, and $D_L$) and the fifth 
component of the gauge field, $A_5^i$, the fifth dimension $y$ is
compactified on an $S^1/Z_2$ orbifold ($Z_2:y \to -y$).
The KK decompositions of the 5D fields are:
\bea
H(x,y ) & = & {1 \over {\sqrt{2 \pi R}}} \left\{ H^{(0)}(x) + \sqrt{2} 
\sum_{n=1}^{\infty}  H^{(n)}(x) \cos(ny/R)\right\}\nonumber\\
A_{i\mu} (x,y ) & = & {1 \over {\sqrt{2 \pi R}}} 
\left\{ A_{i \mu}^{(0)}(x) + \sqrt{2} 
\sum_{n=1}^{\infty}  A_{i \mu}^{(n)}(x) 
 \cos(ny/R)\right\}\nonumber\\
A_{i 5}(x,y ) & = & {1 \over {\sqrt{ \pi R}}} 
\sum_{n=1}^{\infty} A_{i 5}^{(n)}(x) \sin(ny/R)\nonumber\\
Q(x,y )& = & {1 \over {\sqrt{2 \pi R}}} \left\{ Q_L^{(0)}(x) + \sqrt{2} 
\sum_{n=1}^{\infty}  \left[Q_L^{(n)}(x) 
 \cos(ny/R) + Q_R^{(n)}(x) \sin(ny/R)\right]\right\}\nonumber\\
U(x,y )& = & {1 \over {\sqrt{2 \pi R}}} \left\{ U_R^{(0)}(x) + \sqrt{2} 
\sum_{n=1}^{\infty}  \left[U_R^{(n)}(x) 
  \cos(ny/R) + U_L^{(n)}(x) \sin(ny/R)\right]\right\}\nonumber\\
D(x,y )& = & {1 \over {\sqrt{2 \pi R}}} \left\{ D_R^{(0)}(x) + \sqrt{2} 
\sum_{n=1}^{\infty}  \left[D_R^{(n)}(x) 
  \cos(ny/R) + D_L^{(n)}(x) \sin(ny/R)\right]\right\}
\eea
Here the factor of $\sqrt{2}$  is due to the different normalizations of the 
zero and higher modes in the KK tower; it would not have been there if we
run the sum over both positive and negative values of the KK number $n$.  
The fields which are even under the $Z_2$ orbifold symmetry have 
zero modes, and they correspond to the SM particles in usual four dimensions. 
Fields which are odd under $Z_2$ do not have zero modes and hence are absent
in the SM spectrum.

Using the KK expansions of the 5D fields and integrating out 
the 5D Lagrangian over the extra dimension $y$, 
the effective 4D Lagrangian is obtained. Apart from the usual mass term
coming from the vacuum expectation value of the zero-mode Higgs, 
KK excitations also receive masses from the
kinetic energy term in the 5D Lagrangian. The mass of the $n$-th level KK
particle, where $n$ is the KK excitation number that quantises the momentum 
along the extra dimension $y$, is given by $m_n^{KK} =\sqrt{m_0^2 + m_n^2}$ 
where $m_0$ is the zero mode mass and $m_n = n/R$.
Thus the KK spectrum at each
excitation level is nearly degenerate except for the heavy 
SM particles ($t,W,Z,h$). This degeneracy is removed 
by the radiative corrections of KK mode masses \cite{cms1} which play
an important role in collider phenomenology. However, this has only
a negligible effect on our results, and so we can take the KK excitations of 
all the light particles to be degenerate.   
 The couplings $g_5^i$, $\lambda_5^U$ and $\lambda_5^D$ 
are dimensionful and they have
to be rescaled as $g^i=g_5^i/\sqrt{2 \pi R}$, 
$\lambda^U=\lambda_5^U/\sqrt{2 \pi R}$ and 
$\lambda^D=\lambda_5^D/\sqrt{2 \pi R}$  to obtain 
the proper dimensionless SM couplings.

 The zero mode and the KK Higgs doublets 
can be written as
\be
\pmatrix{\phi^{(0)+} \cr {1 \over \sqrt{2}}(v+h^{(0)}+i \chi^{(0)})},
 {\quad} \pmatrix{\phi^{(n)+} \cr {1 \over \sqrt{2}}(h^{(n)}+i \chi^{(n)})}.
\ee
Here $h^{(n)}$s are neutral Higgs KK excitations. 
The charged scalars $\phi^{\pm (n)}$
 combining with the $W^{\pm 5(n)}$ form 
longitudinal components of the $W^{\pm (n)}_\mu$.
 The orthogonal combinations yield physical charged Higgs KK tower.
Goldstone KK modes for $W^{\pm (n)}$ are
\be
G^{\pm (n)}={{m_n W^{\pm (n)}_5 \pm i m_W \phi^{\pm (n)}} 
\over {\sqrt{m_n^2 + m_W^2}}}
\ee
and  the physical charged Higgs KK tower is
\be
H^{\pm (n)}={{m_n \phi^{\pm  (n)} \pm i m_W W^{\pm (n)}_5} 
\over {\sqrt{m_n^2 + m_W^2}}}.
\ee
 Similarly, the $\chi^{(n)}$ together with 
the $Z^{5(n)}$ generate additional physical neutral Higgs tower and 
longitudinal components of the $Z_\mu^{(n)}$. 
However, they do not contribute
 to our study and have consequently not been discussed further. 
In the unitary gauge which we will use for our calculation, 
Goldstone KK modes are eaten up by the longitudinal 
 parts of the gauge boson KK modes. The fields $W_\mu^{\pm(n)}$ and 
$H^{\pm (n)}$ have the same mass as $m_{W,n}=\sqrt{m_n^2 + m_W^2}$, 
since we are neglecting the loop corrections to the KK modes.

Rotating the quark fields to the mass eigenbasis from the weak eigenbasis
is similar to that in the SM and leads to a universal CKM matrix, same for
all KK levels. Furthermore, one generally performs a chiral transformation to
get the 4D mass terms with the correct sign:
\be
\pmatrix{U^{(n)} \cr Q^{(n)}}=\pmatrix {-\gamma_5 \cos{\alpha_n} & \sin{\alpha_n}
\cr  \gamma_5 \sin{\alpha_n} & \cos{\alpha_n}} 
\pmatrix{U^{(n)'} \cr Q^{(n)'}}
\ee
with the mixing angle $\tan {2 \alpha_n}=m_q/m_n$. Obviously, the mixing angles
can be neglected for all quarks except the top.

 For $\bbbar$ mixing, we need the vertices involving one 
zero mode and two non zero KK modes, $W^{+(n)} {\bar d} u_i^{(n)}$, 
$W^{-(n)} {\bar u}_i^{(n)} b$, $H^{+(n)} {\bar d} u_i^{(n)}$ and  
$H^{-(n)} {\bar u}_i^{(n)} b$, to calculate the relevant box diagrams in UED.
 The vertices $H^{+(n)} {\bar d} u_i^{(n)}$ and
$H^{-(n)} {\bar u}_i^{(n)} b$ have two parts, one coming from 
 $\phi^{\pm  (n)}$ interactions, while other part from $W_5^{\pm (n)}$.
All these relevant vertices are obtained from the four dimensional Lagrangian
$\int_{-\pi R}^{\pi R}{\cal L}_5 (x,y) dy$.
The 5D integrations  ${1 \over {(2 \pi R)^{3/2}}}\int_{-\pi R}^{\pi R}
2\cos^2(n y/R) d y$ and  ${1 \over {(2 \pi R)^{3/2}}}\int_{-\pi R}^{\pi R}
2\sin^2(n y/R) d y$ are just $1 \over \sqrt{2 \pi R}$, and combining them 
with $g_5^i$'s we get just the ordinary 4D gauge couplings $g^i$. 
Thus these vertices are exactly identical to the SM ones in weak basis.
 The only point to
note is that the mass terms appearing in the Yukawa couplings of the
 $\phi^{\pm  (n)}$ (relevant for charged Higgs KK mode interactions) 
are the zero-mode and not the excited level masses of the
corresponding quarks. 
The box diagrams relevant for $\bbbar$ mixing in UED
 are shown in Fig.\ 1, to which one must add the crossed diagrams with 
intermediate boson and quark lines interchanged.
In the case of SM, the box diagram 
is mediated only by the exchange of the $W$ boson in the unitary gauge, while 
 in the UED case, the exchange of the KK modes of 
the charged Higgs will give extra diagrams in
 addition to that by the KK excitations of $W$.
\begin{figure}[h]
{\begin{center}
\begin{picture}(460,65)(0,0)
\ArrowLine(0,50)(25,50)
\Text(13,53)[b]{$ b$}
\Text(13,-3)[t]{$\bar d$}
\Text(16,25)[c]{$u_i^{(n)}$}
\Text(85,25)[c]{$u_j^{(n)} $}
\Text(55,-3)[t]{$W^{(n)}$}
\Text(55,53)[b]{$W^{(n)}$}
\Text(87,53)[b]{$ d$}
\Text(87,-3)[t]{$\bar b$}
\ArrowLine(25,0)(0,0)
\ArrowLine(25,50)(25,0)
\ArrowLine(27,50)(27,0)
\ArrowLine(75,0)(75,50)
\ArrowLine(73,0)(73,50)
\ArrowLine(100,0)(75,0)
\ArrowLine(75,50)(100,50)
\Photon(25,48)(75,48){1}{8}
\Photon(25,50)(75,50){1}{8}
\Photon(25,0)(75,0){1}{8}
\Photon(25,2)(75,2){1}{8}
\ArrowLine(120,50)(145,50)
\Text(133,53)[b]{$ b$}
\Text(133,-3)[t]{$\bar d$}
\Text(136,25)[c]{$t^{(n)}$}
\Text(206,25)[c]{$t^{(n)}$}
\Text(180,-3)[t]{$H^{(n)}$}
\Text(180,53)[b]{$H^{(n)}$}
\Text(207,53)[b]{$ d$}
\Text(207,-3)[t]{$\bar b$}
\ArrowLine(145,0)(120,0)
\ArrowLine(145,50)(145,0)
\ArrowLine(147,50)(147,0)
\ArrowLine(195,0)(195,50)
\ArrowLine(193,0)(193,50)
\ArrowLine(220,0)(195,0)
\ArrowLine(195,50)(220,50)
\DashLine(145,50)(195,50){3}
\DashLine(145,48)(195,48){3}
\DashLine(145,0)(195,0){3}
\DashLine(145,2)(195,2){3}
\ArrowLine(240,50)(265,50)
\Text(253,53)[b]{$ b$}
\Text(253,-3)[t]{$\bar d$}
\Text(256,25)[c]{$t^{(n)}$}
\Text(326,25)[c]{$t^{(n)}$}
\Text(300,-3)[t]{$H^{(n)}$}
\Text(300,53)[b]{$W^{(n)}$}
\Text(327,53)[b]{$ d$}
\Text(327,-3)[t]{$\bar b$}
\ArrowLine(265,0)(240,0)
\ArrowLine(265,50)(265,0)
 \ArrowLine(267,50)(267,0)
\ArrowLine(315,0)(315,50)
\ArrowLine(313,0)(313,50)
\ArrowLine(340,0)(315,0)
\ArrowLine(315,50)(340,50)
\Photon(265,50)(315,50){1}{8}
\Photon(265,48)(315,48){1}{8}
\DashLine(265,0)(315,0){3}
\DashLine(265,2)(315,2){3}
\ArrowLine(360,50)(385,50)
\Text(373,53)[b]{$ b$}
\Text(373,-3)[t]{$\bar d$}
\Text(376,25)[c]{$t^{(n)}$}
\Text(446,25)[c]{$t^{(n)}$}
\Text(420,-3)[t]{$W^{(n)}$}
\Text(420,53)[b]{$H^{(n)}$}
\Text(447,53)[b]{$ d$}
\Text(447,-3)[t]{$\bar b$}
\ArrowLine(385,0)(360,0)
\ArrowLine(385,50)(385,0)
 \ArrowLine(387,50)(387,0)
\ArrowLine(435,0)(435,50)
\ArrowLine(433,0)(433,50)
\ArrowLine(435,0)(460,0)
\ArrowLine(460,50)(435,50)
\DashLine(385,50)(435,50){3}
\DashLine(385,48)(435,48){3}
\Photon(385,0)(435,0) {1}{8}
\Photon(385,2)(435,2){1} {8}
\end{picture}
\end{center}}
      \caption{\em UED contributions to the box diagrams for $\bbbar$ mixing.} 
	\label{fig:box_bbbar}
\end{figure}
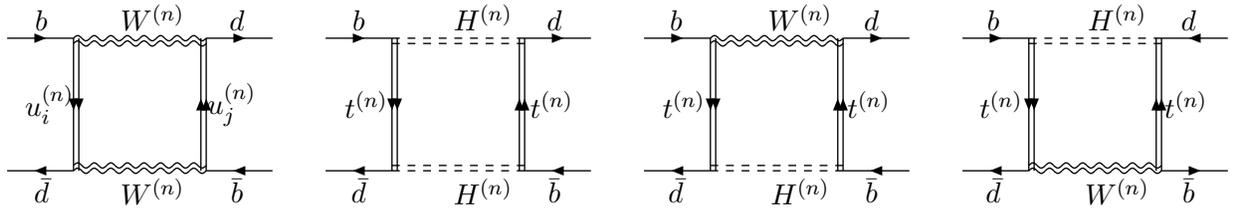

The UED contributions to the effective Hamiltonian for $\Delta B=2$ 
transitions responsible for $\bbbar$ mixing, which come from the box 
diagrams shown in Fig.\ 1, are 
\be
{\cal H}_{\rm eff}^{UED} =  {{G_F^2  \lambda_t^2} \over {4 \pi^2}} 
\left({m_W^4 \over {m_{W,n}^2}}\right)
 \left({\bar d} \gamma_\mu {{(1-\gamma_5)} \over 2}b\right)^2 
 \sum_{n=1}^\infty
\left[S(X_t,X_t)+
 S(X_u,X_u) - 2 S(X_t,X_u) \right]
    \label{h-ued}
\ee
 where
\be
S(X_i,X_j)=S^{WW}(X_i,X_j) + S^{HW}(X_i,X_j) + S^{HH}(X_i,X_j)
\ee
with  $X_i=m_{i,n}^2/ m_{W,n}^2$ and 
$\lambda_i = V_{ib} V_{id}^*$.
The functions $S^{WW}(X_i,X_j)$ come from the box diagram mediated by
two excited $W$s and is of the same form as in the SM \cite{gabrielli}
with the appropriate modification of masses:
\be
S^{WW}(X_i, X_j)=X_i X_j {\Big\{} {\Big[} {1 \over 4} + {3 \over 2} {1 \over
 {1 - X_i}} -{3 \over 4} {1 \over
 {(1 - X_i)^2}} {\Big]} {\ln{X_i} \over {X_i -X_j}}+(X_i \leftrightarrow X_j)
 -{3 \over 4} {1 \over
 {(1 - X_i)(1-X_j)}}{\Big\}}.
\ee
$S^{HH}(X_i,X_j)$ and  $S^{HW}(X_i,X_j)$ are the contributions coming from
the charged Higgs KK box and the mixed boxes 
(with one $W^{(n)}$ and one $H^{(n)}$) respectively:
\bea
S^{HW}(X_i,X_j) & = & {x_n \over {(1+x_n)^2}} (1-x_i) (1-x_j) \left[
{F_1(X_i,X_j) \over 2} - 2 F_2(X_i,X_j)\right]\\
S^{HH}(X_i,X_j) & = & {1 \over { 4 (1+x_n)^2}} (1 + x_i x_n) (1+ x_j x_n) 
 F_1(X_i,X_j)
\eea
with $x_i=m_i^2/m_W^2$ and $x_n=m_n^2/m_W^2$.
 The functional forms of $F_1(X_i,X_j)$ and $F_2(X_i,X_j)$ are given by,
\bea
F_1(X_i,X_j)  & = & {{[X_j^2 (1-X_i)^2 \log{X_j}-X_i^2 (1-X_j)^2 \log{X_i}
 + (1-X_j) (X_j-X_i) (1-X_i)]} \over {(1-X_i)^2 (1-X_j)^2 (X_j - X_i)}}\\
F_2(X_i,X_j)  & = & {{[X_j (1-X_i)^2 \log{X_j}-X_i (1-X_j)^2 \log{X_i}
 + (1-X_j) (X_j-X_i) (1-X_i)]} \over {(1-X_i)^2 (1-X_j)^2 (X_j - X_i)}}
\eea
In the limit $X_i = X_j$, the expressions $S^{WW}(X_i,X_j)$, $F_1(X_i,X_j)$ and
 $F_2(X_i,X_j)$  become
\bea
S^{WW}(X_i,X_i) & = & {{X_i(4-15 X_i + 12 X_i^2 -X_i^3 -6 X_i^2 \ln{X_i})}
\over {4 (1-X_i)^3}}\\
F_1(X_i,X_i) & = & {{1-X_i^2 + 2 X_i \log{X_i}} \over {(1-X_i)^3}}\nonumber\\
F_2(X_i, X_i) & = & {{2-2 X_i + (1+X_i) \log{X_i}} \over {(1-X_i)^3}}\nonumber
\eea
In eq.\ (\ref{h-ued}), we have considered box diagrams with all combinations 
of the three excited up type quarks ($i,j=u,c,t$). The terms 
are then rearranged by eliminating $\lambda_u$ in favour of $\lambda_c$ and
 $\lambda_t$  using the unitarity relation and thus the GIM mechanism 
of the SM is restored. 
The terms containing $\lambda_c^2$ and $\lambda_c \lambda_t$
are $[S(X_c,X_c)
 +S(X_u,X_u) - 2 S(X_c,X_u)]$ and
$ [ 2 S(X_c, X_t) - 2 S(X_u,X_t) + 2 S(X_u, X_u)
 -  2 S(X_c, X_u)]$ respectively and vanish as
$X_c \simeq X_u \simeq { m_n^2 /m_{W,n}^2}$.

The UED contribution to the mass difference for the $\b$
mesons ($\Delta m_B$) is given by
\be
(\Delta m_B)^{UED}  =  {{\vert \langle \b \vert
{\cal H}_{\rm eff}^{UED} \vert \bbar \rangle \vert} \over m_{\b}} .
\ee
The matrix element for $\langle  \b \vert ({\bar d} \gamma_\mu (1-\gamma_5) b)
({\bar d} \gamma^\mu (1-\gamma_5) b) \vert {\bbar} \rangle$ is
calculated in the usual vacuum insertion approximation, and we have
\be
\langle
\b\vert ({\bar d} \gamma_\mu (1-\gamma_5) b)
({\bar d} \gamma^\mu (1-\gamma_5) b) \vert \bbar \rangle=
 {8 \over 3}B_B f_B^2 m_B^2
\ee
which gives,
\be
(\Delta m_B)^{UED}  =  {G_F^2 \over {6 \pi^2}}\sum_{n=1}^\infty
 {m_W^4 \over {m_{W,n}^2}}
 \vert \lambda_t \vert^2
 [S(X_t,X_t)+
 S(X_u,X_u) - 2 S(X_t,X_u)]
B_B f_B^2 m_B \eta_B,
\ee
where the bag factor, $B_B$, is
 introduced to parametrize all possible deviations from the vacuum saturation
 approximation. The quantity $B_Bf_B^2$ has been evaluated from QCD studies
on lattice.  The next-to-leading order (NLO) short distance 
 QCD correction is given by $\eta_B$.  

Let us also comment on the case when one has more than one UED. The electroweak
observables are known be convergent for one UED, but diverges when the number
of UEDs is two or more \cite{zbb}. The SU(3) gauge coupling also becomes 
nonperturbative at high scales for two UEDs, so one has to use a cut-off
scale $M_s$ in the multi-TeV range upto which such perturbative calculations
make sense. This means a natural truncation of the KK mode sum at $n=n_s$
($n_s\sim M_s R$) where $R$ is the common compactification radius. The effect
mainly arises due to a crowding of KK states for more than one UED.  
However, it is gratifying to note that the lower limit on $1/R$ is fairly
insensitive to the exact value of $n_s$ since the terms which do not decouple
vanish due to GIM cancellation.

Now we shall comment on the reliability of the parameters we need
 for $\bbbar$ mixing. The short distance QCD corrections $\eta_B$
 are well determined \cite{qcdcorr}, while the long distance
 corrections are estimated to be small, unlike in the case of $K$-${\bar
 K}$ mixing.  Major uncertainties come from quantities like  
 $V_{td}$, $f_B$ and $B_B$.  We have used the value of $f_B \sqrt{B_B}$
  at a scale of order $m_b$ obtained by UKQCD collaboration \cite{ukqcd} in
 quenched lattice calculation. In the widely used generalised Wolfenstein 
parametrisation of the
 CKM matrix, the CKM matrix elements are expressed 
in terms of the four parameters
 $\lambda$, $A$, $\bar \rho$ and $\bar \eta$.  
Of these, $\lambda$ and A are presently known 
at $1\%$ and $5\%$ levels respectively, 
whereas $\bar \rho$ and $\bar \eta$ are 
the least known CKM parameters.  The uncertainty 
in $V_{td}$ is  solely due to the broad 
allowed range in the $({\bar \rho},{\bar \eta})$ plane. However, one should 
note that since $\Delta m_B$ is affected by UED, the SM fit values of
quantities like $V_{td}$, mainly determined from $\bbbar$ mass difference,
may not be used any further.  
\begin{table}
\begin{center}
\begin{tabular}{|c|c|}
\hline
 & \\[-2ex]
Parameter &  $1 \sigma$ level\\
  & \\[-2ex]
 \hline
$\lambda$ & $0.221\pm 0.002$\\
  & \\[-2ex]
$A$ & $0.83\pm 0.02$\\
  & \\[-2ex]
$\bar \rho$ & $0.151\pm 0.057$\\
  & \\[-2ex]
$\bar \eta$ & $0.369\pm 0.032$\\
  & \\[-2ex]
$|V_{td}| (\times 10^3)$ & $8.36\pm 0.55$\\
  & \\[-2ex]
$f_B \sqrt{B_B}$ & $(235{{+33} \atop {-41}})~~ {\rm MeV}$\\
 \hline
\end{tabular}
\end{center}
  \caption{\em The ranges of relevant 
parameters allowed by UUT at $1 \sigma$ level from \cite{buras2}.} 
\end{table}
  
The models with UED do not have any new local operators beyond those 
already present in SM. In UED, the flavor changing transitions and CP 
 violation are solely governed by the CKM matrix. Furthermore, these 
 models do not contribute to inclusive and exclusive $K$ and $B$ decays 
 due to the absence of KK number violating interactions. 
These properties specify the so-called universal unitarity triangle
(UUT) scenario \cite{buras1}, which 
 has been constructed from
$\vert V_{ub}/V_{cb} \vert$, $\Delta m_B/\Delta m_{B_s}$ and $\sin{2 \beta}$ 
 extracted from the CP asymmetry in $B^0 \to \Psi K_s$ where all dependence on
 $1/R$ cancel out. 
We have performed a complete analysis for $\Delta m_B$ in UED by varying all 
 parameters in their UUT-allowed range, which is specified at the 
 $1 \sigma$ interval in Table 1 \cite{buras2}. 
 The result is shown in  Fig.\ 2 where we have plotted the points
 in  the $1/R-f_B \sqrt{B_B}$ plane compatible with the experimental 
 prediction of $\Delta m_B$ at $1 \sigma$ level.
The lower bound on the compactification scale comes out to be at
165 GeV from the analysis. This is compatible with bounds coming from
other processes, but definitely not better.

\begin{figure}
\begin{center}
\hspace{-1in}
\centerline{\hspace*{3em}
\epsfxsize=14cm\epsfysize=14cm
                     \epsfbox{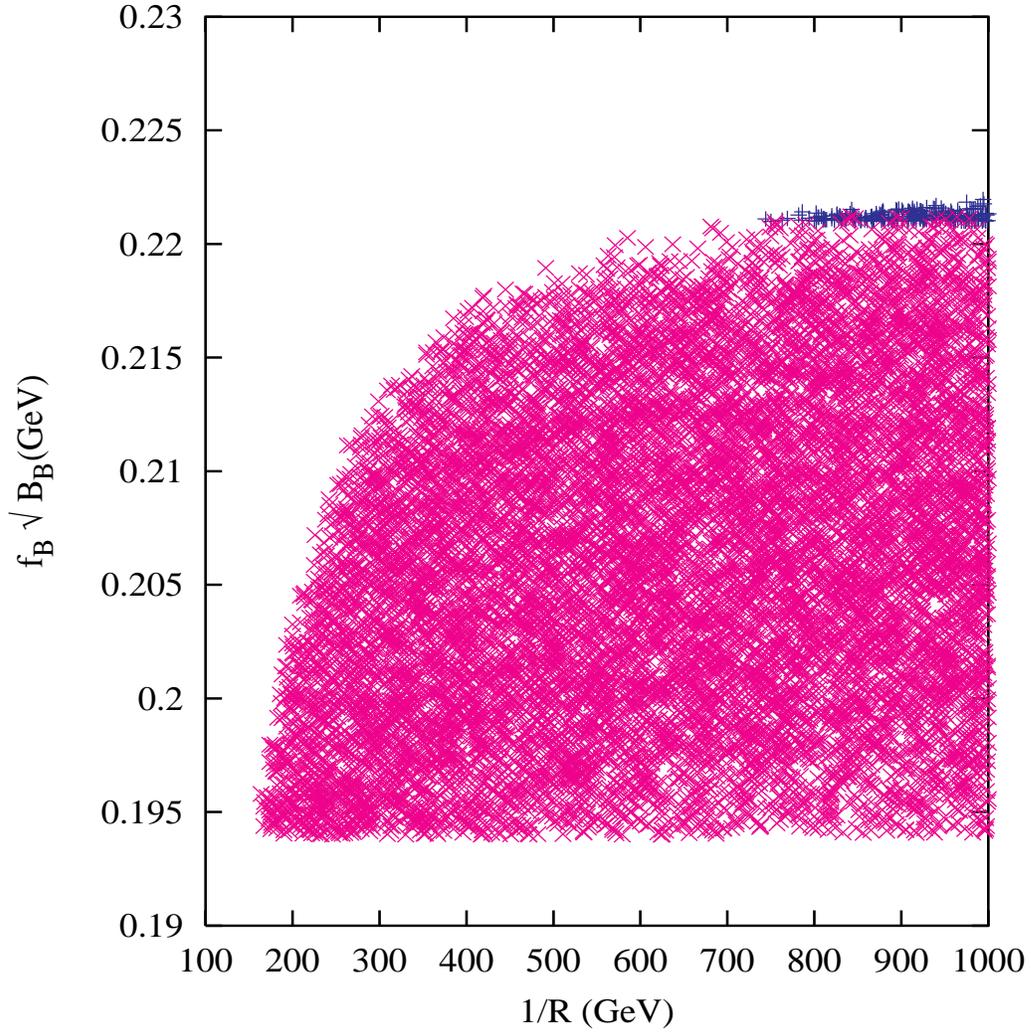}
}
\end{center}
  \caption{\em The region in $1/R-f_B \sqrt{B_B}$ plane allowed by
 the experimental value of $\Delta m_B$ at $1\sigma$ level with the range of the 
 CKM parameters allowed by UUT at $1 \sigma$ level. 
The dark region at the top corresponds to the case when the error on $f_B
 \sqrt{B_B}$ is reduced to one-third of its present value.} 
\end{figure}

Now let us comment on possible future theoretical and experimental 
improvements which may modify this bound drastically. Note that the full
range of the Wolfenstein parameters as mentioned in Table 1 is strongly 
curtailed by the allowed range of $|V_{td}|$, and it is unforeseen in near
future to have such an improvement in these four parameters as to put a 
better constraint on $|V_{td}|$ than that coming from $\bbbar$ mixing. 
Anyway, the bound does not change much for a marginal improvement of
$|V_{td}|$. The case of $f_B\sqrt{B_B}$ is, however, different.
The obtained lower bound is strongly
sensitive to the value of  $f_B \sqrt{B_B}$ as shown in Fig.\ 2.
 If in the future, the error in $f_B \sqrt{B_B}$ is reduced to $1/3$
of its present value, keeping the central value fixed, 
 the lower bound on $1/R$ will be pushed up to 740 GeV.
(Only the dark region in Fig.\ 2 will remain allowed.)
If we use the CKMfitter \cite{ckmfitter} value for this quantity, {\em viz.}, 
$230\pm 28\pm 28$ MeV, the lower bound is close to the present value, but an
improvement by a factor of 3 will only push up the bound to 400 GeV.

In the case of two UEDs, the lower bound on the compactification scale 
(assumed same for both the dimensions)
 is estimated to be about $280$ GeV for $n_s = 5$. This is fairly insensitive 
to the exact value of $n_s$ since the contributions from higher $n$ states
are decoupling in nature due to GIM cancellation.
This bound shots up to 1 TeV if the error bar on $f_B\sqrt{B_B}$ is again
reduced threefold.

It is interesting to note that the UED contribution is always
 positive, increasing the value of $\Delta m_B$ from its SM value.
Thus, if the lower bound on $f_B\sqrt{B_B}$ goes past 222 MeV, the UED 
models will be ruled out or at least $1/R$ will be pushed up to the multi-TeV
range.
The other side of the coin is that if UED has to contribute in a non-trivial
way to $\bbbar$ mixing, lowest-lying KK excitations are going to be detected
hopefully at Tevatron run II, and definitely at the LHC.
  
Let  us now discuss the related process $K$-${\bar K}$ mixing induced by box 
 diagrams in the context of UED. 
This process is governed by $\Delta S=2$ effective Hamiltonian.
 The contribution from the KK  excitations of UED is proportional to
 $\{\lambda_t^2[S(X_t,X_t) + S(X_u,X_u) -2 S(X_t,X_u)] + \lambda_c^2
 [S(X_c,X_t) + S(X_u,X_u) -2 S(X_c,X_u)]+\lambda_c \lambda_t
 [S(X_u,X_u) - S(X_u,X_t) +  S(X_c,X_t)-S(X_c,X_u)]\}$. 
Here $\vert \lambda_t \vert$  (relevant for $\Delta S=2$ interactions)
is suppressed by order $\lambda^4$ ($\lambda \simeq 0.22$ being 
the expansion parameter of the CKM matrix) compared to $\vert \lambda_c \vert$.
In the context of CP violation, the term with  $\lambda_t^2$ may be
 important as $V_{td}$ only contain the CKM phase at ${\cal O}(\lambda^4)$.
The terms proportional to  $\lambda_c^2$ and $\lambda_c \lambda_t$
almost vanish as $X_u \simeq X_c \simeq {m_n^2 /m_{W,n}^2}$.  
  Furthermore, in $K$-${\bar K}$ mixing
 there are large long distance contributions in which the intermediate 
states in the transition $K\to {\bar K}$ are mesons instead of up type quarks
 and bosons. The short distance corrections are  
not well under control even at NLO 
due to the renormalisation scale ambiguity.
 Furthermore, there is still a large uncertainty in the determination of the 
 bag factor $B_K$  for the $K$ system.  
Thus the study of $K$-${\bar K}$ mixing in
the light of UED models is not promising.

To summarize, we have studied the $\bbbar$ mixing in a UED model.
We find lower limit for the compactification scale (165 GeV), which is
close to the limits from other processes.
In addition, we note that even modest theoretical improvements
will have a considerable effect on the bound.
With the error on $f_B\sqrt{B_B}$ reduced to one third, the bound is pushed up
to 740 GeV.
Thus, $\bbbar$ mixing may provide us with a useful tool to discover or strictly
constrain the UED models. Unfortunately, the same cannot be said for the
$K$ system.

{\em Note added.} After this work was completed, a paper came to the 
archive \cite{buras-sw} where the authors have discussed the same effect
in the context of UED and pointed out an error in our calculation.
Their results are in agreement with our corrected calculation.

\begin{flushleft} {\bf Acknowledgements} \end{flushleft}

\noindent
 The authors thank K. Agashe, A.J. Buras, M. Spranger and A. Weiler
 for useful discussions and comments.
They also thank the organisers of WHEPP-7 in HRI, Allahabad, where
this work started.  DC and KH thank the Academy of Finland
(project number 48787) for financial support.
AK's work has been supported by the BRNS grant 2000/37/10/BRNS of DAE,
Govt.\ of India, the grant F.10-14/2001 (SR-I) of UGC, India, and
by the fellowship of the Alexander von Humboldt Foundation.

\newcommand{\plb}[3]{{Phys. Lett.} {\bf B#1} (#3) #2}                  %
\newcommand{\prl}[3]{Phys. Rev. Lett. {\bf #1} (#3) #2 }        %
\newcommand{\rmp}[3]{Rev. Mod.  Phys. {\bf #1} (#3) #2}             %
\newcommand{\prep}[3]{Phys. Rep. {\bf #1} (#3) #2}                   %
\newcommand{\rpp}[3]{Rep. Prog. Phys. {\bf #1} (#3) #2}             %
\newcommand{\prd}[3]{Phys. Rev. {\bf D#1} (#3) #2}                    %
\newcommand{\np}[3]{Nucl. Phys. {\bf B#1} (#3) #2}                     %
\newcommand{\npbps}[3]{Nucl. Phys. B (Proc. Suppl.)
           {\bf #1} (#3) #2}                                           %
\newcommand{\sci}[3]{Science {\bf #1} #2 (#3)}                 %
\newcommand{\zp}[3]{Z.~Phys. C{\bf#1} #2 (#3)}  
\newcommand{\epj}[3]{Eur. Phys. J. {\bf C#1} #2 (#3)} 
\newcommand{\mpla}[3]{Mod. Phys. Lett. {\bf A#1} #2 (#3)}             %
 \newcommand{\apj}[3]{ Astrophys. J.\/ {\bf #1} #2 (#3)}       %
\newcommand{\jhep}[3]{{Jour. High Energy Phys.\/} {\bf #1}:#2 (#3)}%
\newcommand{\astropp}[3]{Astropart. Phys. {\bf #1} #2 (#3)}            %
\newcommand{\ib}[3]{{ ibid.\/} {\bf #1} #2 (#3)}                    %
\newcommand{\nat}[3]{Nature (London) {\bf #1} #2 (#3)}         %
 \newcommand{\app}[3]{{ Acta Phys. Polon.   B\/}{\bf #1} #2 (#3)}%
\newcommand{\nuovocim}[3]{Nuovo Cim. {\bf C#1} #2 (#3)}         %
\newcommand{\yadfiz}[4]{Yad. Fiz. {\bf #1} #2 (#3);             %
Sov. J. Nucl.  Phys. {\bf #1} #3 (#4)]}               %
\newcommand{\jetp}[6]{{Zh. Eksp. Teor. Fiz.\/} {\bf #1} (#3) #2;
           {JETP } {\bf #4} (#6) #5}%
\newcommand{\philt}[3]{Phil. Trans. Roy. Soc. London A {\bf #1} #2
        (#3)}                                                          %
\newcommand{\hepph}[1]{hep--ph/#1}           %
\newcommand{\hepex}[1]{hep--ex/#1}           %
\newcommand{\astro}[1]{(astro--ph/#1)}         %

\end{document}